\documentclass[10pt,conference]{IEEEtran}

\usepackage{cite}
\usepackage{url}
\usepackage{hyperref}
\usepackage{amsmath,amssymb,amsfonts}
\usepackage{algorithmic}
\usepackage{graphicx}
\usepackage{booktabs}
\usepackage{textcomp}
\usepackage{xcolor}
\def\BibTeX{{\rm B\kern-.05em{\sc i\kern-.025em b}\kern-.08em
    T\kern-.1667em\lower.7ex\hbox{E}\kern-.125emX}}
\begin{document}

\title{Interactive Static Software Performance\\Analysis in the IDE}

\author{
  \IEEEauthorblockN{Aaron Beigelbeck}
  \IEEEauthorblockA{
  TU Wien\\
  Vienna, Austria\\
  aaron.beigelbeck@tuwien.ac.at}
  \and
  \IEEEauthorblockN{Maurício Aniche}
  \IEEEauthorblockA{Delft University of Technology\\
  Delft, The Netherlands\\
  M.FinavaroAniche@tudelft.nl}
  \and
  \IEEEauthorblockN{J\"urgen Cito
  \IEEEauthorblockA{TU Wien\\
  Vienna, Austria\\
  juergen.cito@tuwien.ac.at}
  }
}

\maketitle

\begin{abstract}
Detecting performance issues due to suboptimal code during the development process can be a daunting task, especially when it comes to localizing them after noticing performance degradation after deployment. Static analysis has the potential to provide early feedback on performance problems to developers without having to run profilers with expensive (and often unavailable) performance tests.
We develop a VSCode tool that integrates the static performance analysis results from Infer via code annotations and decorations (surfacing complexity analysis results in context) and side panel views showing details and overviews (enabling explainability of the results). Additionally, we design our system for interactivity to allow for more responsiveness to code changes as they happen. We evaluate the efficacy of our tool by measuring the overhead that the static performance analysis integration introduces in the development workflow.
Further, we report on a case study that illustrates how our system can be used to reason about software performance in the context of a real performance bug in the ElasticSearch open-source project.\\
Demo video: \url{https://www.youtube.com/watch?v=-GqPb_YZMOs}\\
Repository: \url{https://github.com/ipa-lab/vscode-infer-performance}
\end{abstract}

\medskip

\section{Introduction}

Software performance issues are often detected and analyzed after the code has already been deployed. 
Understanding software performance in the process of writing code is challenging, as it is often difficult to reason about the performance ramifications of code changes in an ad-hoc manner.
Performance is often analyzed through profilers, which are dynamic analysis tools that attach performance measurements (e.g., latency or CPU utilization) to artifacts in the code.
However, the proper use of profilers for performance analyses requires executing the code with appropriate workloads, which are difficult to design and introduce delays that obstruct the development workflow. 
Static analysis, on the other hand, is a more lightweight opportunity to provide early feedback to developers.
A contemporary static performance analysis approach, implemented within the tool Infer~\cite{Distefano2019}, is based on an efficient algorithm for parametric worst-case execution time calculation~\cite{Bygde2011}.
It performs a static estimation of the exact cost of each function (\emph{cost} being a proxy for underlying instruction cost in execution runtime), which can be abstracted into runtime bounds as code complexity metrics in Big-$\mathcal{O}$ notation.
This can provide developers with a notion of software performance in familiar notation for asymptotic growth known from complexity analysis attached to their functions (e.g., a function being quadratic in its parameter $size$ would be indicated as $\mathcal{O}(size^2)$).
In its current form, the results reported from this tooling are fraught with perils that are generally known from static analysis tooling~\cite{Barik2016}. Producing analysis results requires developers to start a separate process that involves a full-program build, which then outputs potentially hundreds of results. This interrupts development workflows when working inside the IDE, leads to split-attention effects and cognitive fatigue when attempting to interpret results.

We present a system we developed in VSCode for static performance analysis with Infer\footnote{\url{https://github.com/ipa-lab/vscode-infer-performance}}, that enables interactive reasoning about static performance properties in the code that is integrated into the development workflow~\cite{Sadowski2018}.
We hook our process into the editing process of the IDE and use heuristics to determine whether to display analysis results based on incremental analysis or indicate to the developer to re-execute due to the higher probability that the introduced change has affected performance outcomes.
We then apply lightweight program analysis to attach analysis results to function declarations (shown in Figure \ref{fig:overview}, left), but also show how performance properties evolve through code changes (shown in Figure \ref{fig:overview}, right).
To evaluate the efficacy of our tool, we perform two analyses. We perform experiments on open-source Java projects of varying sizes to provide quantitative insights into the overhead that the static performance analysis integration introduces in the development workflow, showing that our integration and analysis steps are negligible compared to dominating build times.
Further, we report on a case study that illustrates how our system can be used to reason about software performance in the context of a real performance bug in the ElasticSearch open-source project.

\section{Integrated and Interactive Static Performance Analysis}

\begin{figure*}[ht]
\centering
\includegraphics[width=0.95\textwidth]{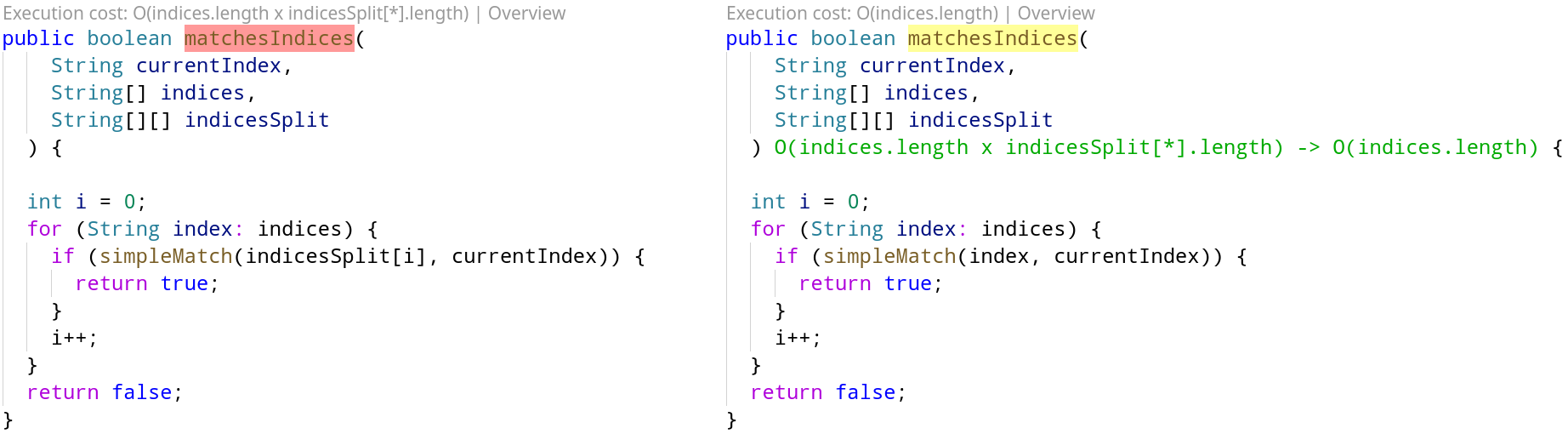}
\caption{Code Annotations in VSCode show performance information as Big-$\mathcal{O}$ as part of the function declaration and appropriate color-coging. Left: A detected performance bug in ElasticSearch from our case study. Right: After introducing the bug-fixing change, we update the analysis result, but also enable traceability by showing how performance properties evolved (in green).\label{matchesIndices_comparison}}
\label{fig:overview}
\end{figure*}

We describe our system structured by our high-level goals \emph{workflow integration}, \emph{enabling interactivity}, and \emph{traceability and explainability}, while interleaving design considerations with implementations for brevity.

\medskip
\textbf{Workflow Integration.}
The way in which static analysis tools present their results is often suboptimal for consumption by software developers~\cite{Barik2016}. Split-attention effects can occur when obtaining analysis results requires initiating a separate program, which is then shown in a separate window from the development environment. This requires the user to context-switch between different interfaces. We propose to integrate this information by directly attaching to the software artifact responsible (i.e., ideally right next to the pieces of code that the analysis output refers to).

To enable this code annotation, we parse the performance result (a larger JSON file containing semi-structured information about the analysis) into an internal structure that maps fully-qualified method names to their particular performance costs. We then incrementally perform the mapping from function declarations in the source code by applying lightweight program analysis to match function declaration in source code files to their corresponding performance result from our internal map.

To visualize the performance result with the declaration, we make use of VSCode's \textsc{CodeLenses}. They provide a way of inserting interactive text above a given line of code (see Figure~\ref{matchesIndices_comparison}). Two \textsc{CodeLenses} are used for each function to provide an increasingly more detailed perspective of the result: one that shows the cost in Big-$\mathcal{O}$ notation and provides a more detailed view when clicked, and another one that leads to an overview of all the functions in the current file and their respective costs.
Color coding is used on the function name, reflecting the severity of the analysis result. The function is seen in green background color if it has a constant cost ($\mathcal{O}(1)$), yellow if linear ($\mathcal{O}(n)$), and red if quadratic or worse ($\mathcal{O}(n^k)$). Hovering over the function name also shows the exact cost without having to open the detail view.

To avoid the potential problem of cognitive overload, we only display the most relevant information directly where the code is located, and a more detailed view can be opened manually. The detail view complements the Big-$\mathcal{O}$ information by also showing its exact cost, as computed by parametric worst-case execution time in Infer~\cite{Bygde2011}. It further includes an evolutionary perspective of how changes within the file have affected performance, which we explain in more detail below.

\medskip
\textbf{Enabling Interactivity. }
Another major goal was to make our system as interactive as possible. Since function performance can easily change on the basis of one seemingly simple code change, it would be optimal to always have up-to-date performance information by re-analyzing the code with every change. However, this is not effectively possible due to limitations that do not enable us to only perform partial or incremental analysis. The underlying tool, Infer, has to capture information from the compilation process itself, on which it can then perform the analysis. Having to re-compile all necessary files for every code change would introduce too much of an interruption in the development process for projects that go beyond just a couple of files, rendering the option of continuous re-execution unviable.

To still enable interactive performance analysis, we implemented a heuristic that computes a significance score of a code change with respect to the probability of a change introducing a performance shift.
We perform source code differencing on every edit to determine performance-sensitive changes, such as added, removed, or changed loops or function invocations in the function body. The developer then gets informed via the \textsc{CodeLenses} that the performance might have changed, and a re-execution of the analysis could make sense to receive performance data that reflects the current code structure. When possible, we only trigger an incremental build to speed up the entire process.

\medskip
\textbf{Traceability and Explainability. }
We want developers to be able to trace the evolutionary impact of code changes on performance.
Therefore, whenever the performance of a function has significantly changed between code changes, the change in Big-$\mathcal{O}$ cost is shown on the right side of the parameters as an evolution step ($\mathcal{O}(indices.length \times indicesSplit[*].length) \longrightarrow \mathcal{O}(indices.length)$ in Figure~\ref{matchesIndices_comparison}, right). This is complemented with color-coding to make it immediately visible whether the performance got better or worse.
Another problem that static analysis tools often face is that they operate more or less like black boxes. They receive code pieces as input, perform internal operations that are not always fully comprehensible for the developer, and give results that could be better understood if some of the information from the analysis' internal workings would be made accessible to the developer~\cite{Barik2016}. And even if not extracted from the analysis process itself, heuristic approaches that try to reason about the analysis output after the fact could lead to more comprehensive explanations of the underlying problem.
We provide further traces in our detailed view that explains where the calculated cost comes from, i.e., loops and method calls, including their parameters. This gives developers the opportunity to understand why a function has a particular cost, enabling them to reason about potential performance issues and fixes more effectively. Furthermore, instead of just presenting the reasoning behind one specific cost calculation, we try to heuristically detect the code changes that might have led to a change in cost, as explained earlier. These are also shown in the detail view, both for cost changes that have already occurred as for code changes that might lead to a significant cost change when the analysis gets re-executed again, enabling early feedback.

\section{System Overhead Analysis}
\label{sec:overhead}

We measured the time it takes to run Infer on several projects and load the performance data into VSCode with our system to provide quantitative insights into overhead such an integrative process into the development workflow.
\textbf{Experiment Setup. } Measurements were taken on six different projects with varying size: a tiny one ($<$100 LOC) to see whether there is significant overhead even when there is a negligible amount of code to analyze, four medium to large-sized projects (about 20.000-150.000 LOC) to check the viability of the extension for regular day-to-day projects, and a very large project ($\sim$450.000 LOC) to stress test the performance of the system on loading the data into VSCode and performing program analysis for matching and displaying.
We conducted the experiments on a laptop running Manjaro Linux with an 8-core Intel i7 processor and 8 GB of RAM.
The results show that the compilation plus the capturing phases of Infer take by far the longest time. The analysis is usually much faster, and loading the data into the system only took roughly half a second, even on the largest project.
\medskip

\textbf{Full-Build Experiment Results.}
Table~\ref{performance_table} shows an overview (median over five runs) of the quantitative results.
More concretely, compilation plus capturing on the tiny project still took 5 seconds, probably due to the overhead of using a build tool (Gradle, in this case) for such a small code base, whereas the other steps combined took less than 100 ms. The compilation plus capturing times for the larger projects were somewhere between 20 to 90 seconds, linearly depending on the size. Interestingly, the analysis phase also took a larger portion of time for the larger projects than for the smaller ones. On the project with $\sim$50.000 LOC, the analysis took about 1/10 of the time for compilation plus capturing, and on the project, with $\sim$150.000 LOC it was around 1/4.
Since running Infer on very large projects takes a significant amount of time ($\sim$170 seconds for compilation, capturing, and analysis combined on the project with $\sim$450.000 LOC), the system might not be suitable for re-running the analysis multiple times to see the changes in performance. However, if the performance data from Infer is already available and the developer just wants to have their code annotated with it, then loading the data into the extension comes at negligible overhead.

\medskip
\textbf{Incremental Analysis Result. }
Aside from performance measurements related directly to Infer, we also measured the time for the extension to check for potentially significant code changes after a file gets saved and displaying this information to the developer. Since this process is only done incrementally (for one active file at a time), project size is not a variable of interest to our experiment, but rather file size and the number of code chunks as part of the edit operation. After some experimentation, it was clear that this feedback is generated almost instantly in any case, since even on a file with $\sim$5.000 LOC it took about 10 ms on average to detect one potentially significant code change at the end of the file.

It is also noteworthy that Infer supports build tools with incremental compilation. Depending on the build tool, project, and how granular the incremental compilation is in any specific instance, this can potentially speed up the time for compilation plus capturing plus analysis significantly after the first execution.
However, we opted to be conservative in our reporting and only present results on full builds.

\begin{table}
\centering
\caption{The median values of 5 runs each, measuring the runtimes of the individual steps on differently sized projects.\label{performance_table}}
\begin{tabular}{p{1.8cm}p{1cm}p{1.5cm}p{0.8cm}p{1.75cm}}
\toprule
GitHub Project & LOC & Compilation + Capturing & Analysis & Loading into Extension\\
\midrule
hello-world & 64 & 5 s & 46 ms & 10 ms\\
biojava (core) & $\sim$20k & 8 s & 2 s & 42 ms\\
Netflix/Hystrix & $\sim$50k & 29 s & 3 s & 61 ms\\
OpenRefine & $\sim$60k & 21 s & 5 s & 87 ms\\
biojava (full) & $\sim$150k & 87 s & 21 s & 238 ms\\
elasticsearch & $\sim$450k & 127 s & 39 s & 524 ms\\
\bottomrule
\end{tabular}
\end{table}

\section{Case Study}

To illustrate the efficacy of our system in the context of reasoning about software performance in the development workflow, we report on a case study on a performance bug in the search engine ElasticSearch.\footnote{The performance bug is present in ElasticSearch's commit 4a01879 and was inspired by a post in the blog ``Accidentally Quadratic"}
We first briefly explain the context of this performance bug before we showcase how our system can help fix it.

\medskip
\textbf{Case Context.}
Elasticsearch stores indices separated by dashes ($-$) as tokens, \texttt{shard1-shard2-shard3-\dots} (a concrete example could take this form \texttt{tweets-01-01-2020}). 
The method under analysis takes two arguments: \texttt{currentIndex} and \texttt{indices}, with the latter containing patterns that are used to filter for all the concrete indices from our stored ones that correspond to these patterns, and the former being used as the pattern for our final search result. Further on, we refer to the filtered concrete indices simply as \texttt{indices}, not meaning the initial patterns used as the filters.
For the sake of simplicity in this demonstration, we sliced out the part of the code that represents the cause and effect of the performance bug. This replication happens without loss of generality, as we carefully extract the parts of the functionality that fully represents the challenges inherent for reasoning about this performance issue.
Our re-implementation of the \texttt{matchesIndices} function has three arguments, with the third one simply being the \texttt{indices} array where each entry has been split at the dashes, which gives us the individual shards of each index.
The problem now becomes that we are looping through all the \texttt{indices}, which could in the worst case be all the stored indices, and for each one of these indices, we match the \texttt{currentIndex} with every shard, which can also be arbitrarily many, \emph{leading to quadratic cost}.

\medskip
\textbf{Integrated Performance Reasoning.}
Our system was able to detect this issue in the development process and provide an early warning for the developer by properly color-coding the function and warning about the performance cost (as can be seen in the left part of Figure~\ref{matchesIndices_comparison}). With the trace provided in the detail view, the developer can immediately learn that looping through the \texttt{indices} and calling the matching function within the loop leads to the problematic performance cost.

The developer then fixes this issue by matching \texttt{currentIndex} with the complete \texttt{indices} respectively, instead of with each individual shard. This leads to linear instead of quadratic cost since the matching now happens in constant time.
This change in cost was also illustrated by the system, now color-coding the function name yellow (for linear cost), and showing the evolution in performance cost next to the parameters of the function with green coloring, since the cost improved (see right part of  Figure~\ref{matchesIndices_comparison}). If interested in comparing the new detailed cost traces to the previous one, this can be done in the detailed view that includes the cost history of the function.

With this case study, we wanted to illustrate that our system provides an integrated way for developers to get valuable information about the performance of their code in an unobtrusive and intuitive way.

\section{Discussion}

We briefly discuss issues we encountered in the design of our system with respect to uncertainty in static analysis and the overhead of requiring build and compilation steps.

\medskip
\textbf{Analysis Uncertainty.}
One major problem that occurs whenever Infer is unable to analyze some parts of the code is that the cost of all functions in the call chain depending on this code cannot be computed. This means that one line of code that has an unknown cost, regardless of its significance, would be enough to render the whole function incomputable with respect to performance cost, which then also transitively propagates to all other functions in the call chain.
This design choice, albeit conservative, is understandable to avoid compounding errors. If the analysis would ignore the incomputable parts of the code and just output the cost of the computable parts, the result could potentially be inaccurate, which would again propagate to all the dependent functions. However, some heuristic for dealing with certain incomputable constructs, together with the ability for the developer to opt for a more risky configuration (e.g., assuming constant cost for uncomputable entities, for instance), may represent a reasonable trade-off here.

\medskip
\textbf{Interactivity Bottleneck: Build Times.}
A major obstacle to enabling truly interactive performance analysis is the need for rebuilding the code. 
Seamless integration of static analysis tools into the regular workflow of developers is crucial for their adoption~\cite{Sadowski2018}. To enable this kind of smooth integration, we envision designing and partial builds with the distinct purpose of building the base for incremental program analysis in mind to reduce build times.
There is already some related work, specifically for taint analysis, that heavily invests in seamless static analysis techniques to avoid disrupting the developers' workflow~\cite{Do2017}.

\section{Related Work}

The work most related to our tool is \emph{PerformanceHat}, which collects runtime traces in production environments and integrats them as part of the development workflow as into source code artifacts in the Eclipse IDE~\cite{Cito2019}.
Work on scalability of static analysis approaches and their real-world adoption by developers is another popular concern.
Distefano et al.\ describe how advanced static analysis techniques are able to deal with very large industrial code bases and catch important defects~\cite{Distefano2019}.
Cheetah wants to make static analysis more interactive by performing incremental analysis ranging from less to more complex computations~\cite{Do2017}.
MagpieBridge is a generalized approach for static analysis integration into the IDE with the goal of reducing the complexity of the integration process~\cite{Luo2019}.


\section{Conclusion}

We presented a system implemented in VSCode that integrates static performance analysis results into the development workflow. It contextually matches performance properties obtained from the static performance analysis in Infer to code artifacts (function declarations) through lightweight program analysis. It enables interactivity by performing source code differencing on edits to heuristically determine whether re-execution of the analysis is necessary. 
We illustrate the efficacy of our system in a case study about a performance bug in the search engine ElasticSearch. Our system makes it possible to reason about performance issues effectively in the code much earlier and supports performance fixes in the development workflow.
We also show through quantitative experiments that the integration and analysis steps of our system are negligible compared to the build times required for the analysis and discuss potential ways forward to improve interactivity.

\bibliographystyle{IEEEtran}
\bibliography{references}

\end{document}